# Inertial polarization of dielectrics

A.G. Zavodovsky[*]

*Surgut State University,
Energetikov str.,14, 628400, Surgut,
Khanty-Mansiysk Autonomous Okrug, Tyumen region, Russia*

**Abstract**

It was proved that accelerated motion of a linear dielectric causes its polarization. Accelerated translational motion of a dielectric's plate leads to the positive charge of the surface facing the direction of motion. Metal plates of a capacitor were used to register polarized charges on a dielectric's surface. Potential difference between the capacitor plates is proportional to acceleration, when acceleration is constant potential difference grows with the increase of a dielectric's area, of its permittivity and is hardly dependant on a dielectric's thickness.

A dielectric is polarized by an applied electric field. Polarization is a polarized charge on the surface. There are 3 types of polarization: deformational, dipole, polarization of ions. The type of polarization depends on the type of a dielectric. Some anisotropic dielectrics (piezoelectrics) can be polarized without application of the electric field, if a dielectric experiences mechanical deformations. Piezoelectrics include a group of dielectrics (pyroelectrics) which are characterized by a spontaneous polarization without exposure.

The experiments by R.Tolmen and T.Stuart exposed electron-inertia effects 90 years ago [1]. The analysis of the received data showed that inertial forces arising in a conductor generate an external field, kinetics of conduction electrons is described in the same way as in the case when a conductor is placed in an electric field [2]. Consequently we can assume that extraneous forces (inertia) in a dielectric cause polarization. This assumption was proved by an experiment.

The scheme of the experiment is given in Figure 1. The plate of a linear dielectric 1 and metal plates 2 (flat capacitor) were fixed in a small closed copper container 3, which could freely move within a metal cell 4 in directions A and B perpendicularly to the surface of the plate. The capacitor plates were switched to "+" and "—" terminals of the recording device 5 (a self-balancing potentiometer). Flat capacitors with surfaces ranging from 5 to 15 sm$^2$ were used in the research. Thickness d of a dielectric's plate varied from 1.3 to 3.2 mm. Samples under research included linear dielectrics with permittivity ε from 2.0 to 5.5, the quantity was measured by a capacitance method. A special mechanical device was introduced to produce a regulated impulse action on a container with a capacitor, the device induced different acceleration to the capacitor. Acceleration was measured in arbitrary units.

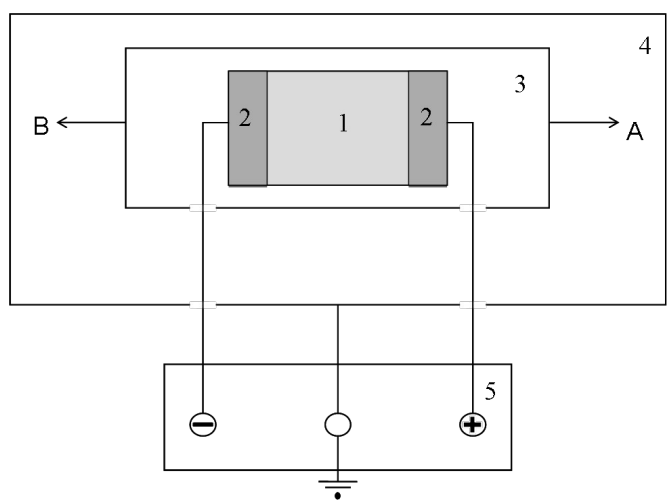

Figure 1: The scheme of the experiment: 1 — a dielectric's plate, 2 — metal plates, 3 – copper container, 4 – cell, 5 – recording device.

---

[*]E-mail: averin117@mail.ru

When the container 3 with an uncharged capacitor inside remained at rest, or moved at a constant velocity in directions A and B relative to cell 4, potential difference $\Delta\varphi_0$ on capacitor plates was zero. In case of accelerated translational motion potentiometer 5 indicated non-zero potential difference $\Delta\varphi_0$ on the capacitor plates, i.e. a polarized charge arose on a dielectric's surface. An experiment discovered that a dielectric's surface facing the direction of motion acquired a positive charge. The direction alteration of a capacitor's motion changes "+" to "—" in a potential difference between capacitor plates. Figure 2 demonstrates the result of impulse action (in directions A and B) on a capacitor.

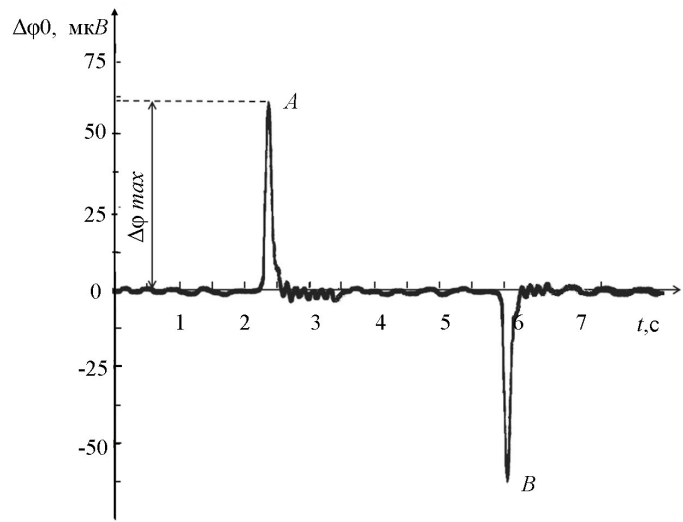

Figure 2: Potential difference $\Delta\varphi_0$ on the capacitor plates as a function of time (duration) of impulse action in directions A and B.

The experimental data were processed to determine the magnitude of a potential difference $\Delta\varphi_{max}$ corresponding to a definite acceleration of a capacitor under certain conditions. An empty container 3 was also under experiment to take into account potential difference of "background" $\Delta\varphi_f$. A real polarization potential difference $\Delta\varphi$ was calculated by extracting "background" $\Delta\varphi_f$ from $\Delta\varphi_{max}$.

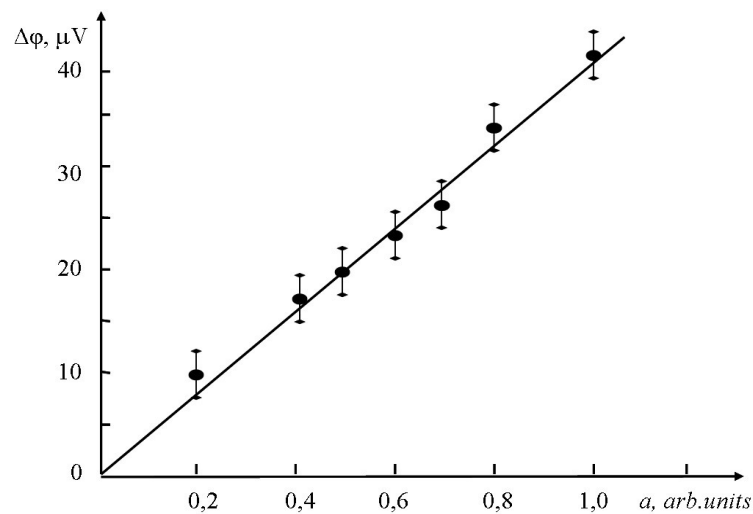

Figure 3: Potential difference $\Delta\varphi$ as a function of a capacitor's acceleration (S=15 sm$^2$, d=1.3 mm, ε=5.5).

The analysis of the received data showed that $\Delta\varphi$ depends on acceleration of a dielectric. The experimental research established the type of this relation. The outcomes are showed in Figure 3. Taking into account the experimental margin of error this type of relation can be referred to as linier. Thus an increase in acceleration produces bigger polarized charge on a dielectric's surface, the alteration of motion direction affects "+" or "—" of a charge.

Experiments were also carried out to determine the relation between polarized potential difference and properties of a dielectric's plate (surface area and thickness). Given constant acceleration potential difference rises with the increase of a dielectric's surface area S. This function is plotted in Figure 4. The experimental research demonstrated: that a dielectrical plate's thickness increases 2.5 times, potential difference is hardly boosted. At given constant acceleration, surface area S and thickness d of a dielectric's plate the study was carried out to establish potential difference $\Delta\varphi$ as a function of permittivity $\varepsilon$. The analysis of the experimental findings shows that under particular experimental conditions potential difference $\Delta\varphi$ grows with an increase of permittivity $\varepsilon$ in a non-linear way.

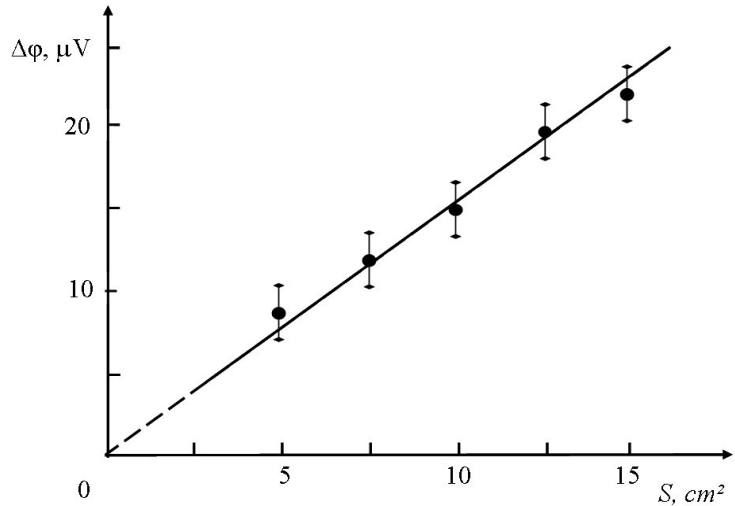

Figure 4: Potential difference $\Delta\varphi$ as a function of a dielectrical plate's surface area, acceleration a=0.5 arb.units (d=1.3 mm, $\varepsilon$=5.5).

A model of electronic polarization accounts for the observed effect. It can be assumed that inertia generated by an accelerated motion of a dielectric displaces electron shells of atoms. The experimental data show that the degree of electron displacement is proportional to acceleration. Larger surface area of a dielectric accumulates bigger polarized charge, thus producing bigger potential recorded. Weak relation between potential and a dielectric's thickness can be presumably accounted by the fact that the observed effect has considerable magnitude and the polarizability affects the quantity of the surface charge. Besides, the inhomogeneity of the field near the edges of the capacitor influences the potential as well.

The paper reveals the data which can be used to invent various technical devices. The experimental findings allowed to work out an acceleration-sensing element [3]. There is an ample opportunity for developing devices aimed at direct transformation of mechanical energy into electrical energy. The research findings contribute to accounting for accumulation of electric charge in dusty clouds. It became known long ago that colliding between each other sand or dust particles can accumulate great electric charge causing thunderstorm discharges in deserts and explosions on sugar and coal-processing enterprises. However the mechanism of charges' emergence has not been still revealed. The publication [4] tested the model stating the following: after dielectric particles' polarization in applied electrical field particles can accumulate big electrical charge as a result of numerous collisions (despite their initial zero charge). The experimental findings confirmed outcomes of this model. But the authors of the publication were not able to determine the source of polarizing electrical field in natural and industrial processes. The ideas of this article give evidence for stating that polarization is produced by the inertial forces generated from an accelerated motion of dielectric's particles.

## References

[1] R.C. Tolman, T.D. Stewart, Phys.Rev. **8**, 97 (1916).

[2] I.M. Tsidilkovsky, The Soros Educational Journal **6,** №9, 87 (2000).

[3] A.G. Zavodovsky, Russian patent for the utility model № 92965. Newsletter №10. (10.04. 2010).

[4] T. Pahtz, H.J. Herrmann & T. Shinbrot, Nature Phys. advance online publication doi: 10.1038/NPHYS16.31 (11 April 2010).